# On the Debye temperature in sigma-phase Fe-V alloys

Jakub Cieslak[1], Michael Reissner[2], Stanislaw M. Dubiel[1*], Benilde F. O. Costa[3] and Walter Steiner[2]

[1]Faculty of Physics and Computer Science, AGH University of Science and Technology, 30-059 Krakow, Poland, [2]Institute of Solid State Physics, Vienna University of Technology, 1040 Wien, Austria, [3]CEMDRX Department of Physics, University of Coimbra, 3000-516 Coimbra, Portugal,

**Abstract**

A series of σ-phase $Fe_{100-x}V_x$ samples with $34.4 \leq x \leq 59.0$ were investigated by neutron and X-ray diffraction and Mössbauer spectroscopy (MS) techniques. The first two methods were used for verification of the transformation from α to σ phase and they also permitted to determine lattice parameters of the unit cell. With MS the Debye temperature, $\Theta_D$, was evaluated from the temperature dependence of the centre shift, <CS>, assuming its entire temperature dependence originates from the second-order Doppler shift. To our best knowledge, it is the first ever-reported study on $\Theta_D$ in σ-FeV alloys. Both lattice parameters i.e. $a$ and $c$ were revealed to linearly increase with $x$. $\Theta_D$ shows, however, a non-monotonic behaviour as a function of composition with its extreme values between ~425 K for $x \approx 40$ and ~600 K for $x \approx 60$. A local maximum of ~525 K was found to exist at $x \approx 43$.

- Corresponding author: dubiel @novell.ftj.agh.edu.pl

## 1. Introduction

Among over 50 examples of the σ-phases (tetragonal unit cell - structure type $D^{14}_{4h}$ $P4_2/mnm$) known to exist in binary alloy systems two viz. σ-FeCr and σ-FeV are exceptional as they exhibit well evidenced magnetic order [1-6]. It is of scientific interest to know if the similarities between the magnetic properties in the two alloy systems with respect to the σ-phase is found also for other physical properties, like dynamical ones. Also of importance is the knowledge of how various physical properties characteristic of the σ-phase depend on alloy composition. In the available literature there is little information on the dynamical properties of the σ-phase, and on the Debye temperature, $\Theta_D$, in particular. To our best knowledge, $\Theta_D$ – values are only known for the σ-phase in the Fe-Cr system, where this phase occurs in a range of about 5% around the equiatomic composition and shows a linear behaviour versus composition [7,8].

In the light of the above given information, it is of interest to verify whether or not similar effects can be found for the σ-FeV phase, which is in fact more appropriate for testing verious characteristics of σ, because the composition range of the σ-phase occurrence in Fe-V is by a factor ~6 larger than the one in the Fe-Cr system [9]. The present paper reports the results found with neutron and X-ray diffraction as well as with Mössbauer spectroscopic techniques for a series of the sigma-phase $Fe_{100-x}V_x$ samples with $34.4 \leq x \leq 59.0$.

## 2. Sample preparation

Master alloys of α-$Fe_{100-x}V_x$ with a nominal composition of vanadium between x = 36 and 60 were prepared by melting appropriate amounts of Fe (99.95% purity) and V (99.5% purity) in an arc furnace under protective argon atmosphere. The ingots received after melting were next solution treated at 1273 K for 72 hours followed by a water quench. The chemical composition was determined on the homogenized samples by electron probe microanalysis as $x = 34.4, 37.6, 39.9, 42.5, 43.6, 46.0, 47.9, 51.3, 55.0, 57.0$ and $59.0$. The transformation into the σ-phase was performed by annealing the ingots at $T_a$ = 973 K for 25 days. The verification of the α- to σ-phase transformation was performed by recording room temperature X-ray and neutron diffraction patterns. It should be noted that from microprobe analysis the σ-phase samples showed some small fraction of V-rich precipitates, most likely in form of carbides. These precipitates were to small to be visible in the neutron diffraction patterns. However, it is known that carbide nucleation often precedes σ-phase nucleation [10]. In addition, one nanocrystalline sample with $x$ = 53.0 was prepared by mechanical alloying of iron and vanadium, both in form of a powder, following the procedure described in Ref. 8. The transformation of such obtained α-phase into the σ-phase was performed by isothermal annealing in argon atmosphere at $T_a$ = 1000°C for 10 hours. The verification of the final phase was carried out by recording room temperature XRD pattern.

### 3. Results and discussion

#### 3.1. Lattice paramaters

The lattice parameters $a$ and $c$ of the unit cell were determined by means of neutron diffraction measurements performed at ILL Grenoble (D1A) on the samples with $x$ = 34.4, 43.6, 47.9 and 59.0, and with X-ray diffraction for $x$ = 39.9, 43.6, 47.9 and 55.0 The diffractograms obtained with $\lambda$=1.91127Å for neutrons and $\lambda$ = 1.54053 Å for X-rays, one example of which is shown in Fig. 1, were analysed by means of Rietveld method (FULLPROF program [11]). There were 22 refined parameters taken into account; 9 of them related to the background and position of the spectrum, next 8 parameters were connected with the scale parameter, line-widths and lattice constants, and 5 parameters were relevant to the Fe/V occupation numbers of the five different lattice sites. In the calculations, the alloy concentration, $x$, was held fixed to the values obtained by the microprobe analysis, while the Fe to V ratio for each of the five lattice sites was a free fitting parameter. To take into account uncertainties in the concentration obtained from the microprobe analysis for each spectrum the full fitting procedure was repeated for two additional compositions: $x+\Delta x$ and $x-\Delta x$ ($\Delta x=0.3$at% being the maximum expected error of the composition determination). The differences between the results of these two calculations were treated as the errors of the particular occupation numbers. The thus obtained values of $a$ and $c$ as well as of the unit cell volume, $V$, are presented as a function of vanadium content, $x$, in Fig. 2. Details concerning the obtained occupation numbers are given in Ref. 12. In first approximation the lattice parameters show a linear dependence on $x$. The increase with $x$ from 34.4 to 59.0, which for $a$ is equal to 0.00508 Å/at % V, for $c$ to 0.00180 Å/at % V and 0.564526 Å$^3$/at % V for the volume of the unit cell, $V$, results in a relative enhancement of $a$ by 1.41%, that of $c$ by 0.095% and that of the unit cell volume by 3.62%. Such changes in the lattice parameters can be, at least qualitatively, understood in terms of the difference in the atomic volume between Fe and V atoms. Off this trend lies the point for the nanocrystalline sample, which may be due to a distorsion of the lattice due to a high ratio between the near-surface atoms to those in the bulk, a charcateristic feature of nanocrystalline samples.

### 3. 2. Debye temperature

$^{57}$Fe Mössbauer spectra were recorded in the temperature interval of 80 – 300 K with a standard spectrometer and a $^{57}$Co/Rh source of the γ-rays. The temperature of the samples, which were placed in a flow cryostat, was kept constant to within ±0.2 K. The shape of the spectra sensitively depends on the sample composition as shown in Fig. 3, and, for a given composition, on temperature – see Fig. 4. The spectra, because of their ill-resolved shape and the underlying complex crystallographic structure (five different crystallographic sites with high (12-15) coordination number and chemically disordered distribution of Fe and V atoms over these sites [12]) were fitted with the hyperfine field distribution (HFD) method. It was assumed, in the fitting procedure, that the hyperfine field is linearly correlated with the isomer shift and with the quadrupole splitting. The former assumption is justified by both

experimentally found evidence for such correlation in binary bcc-alloys of iron alloys including Fe-V [13] as well as on the recent theoretical calculation of the electronic structure for a σ-FeCr compound in the paramagnetic state [14]. According to the latter, changes in the electronic structure due to Cr atoms are in line with the corresponding changes in the α-FeCr alloys. Concerning the correlation between the hyperfine field and the quadrupole splitting, there is weaker evidence for that, though there is also some experimental data in favour of it [15,16]. In addition, the fitting procedure was carried out assuming a ratio of 3:2:1 for the line amplitudes within the sextets. All line widths were assumed to be the same. The Debye temperature was evaluated from the temperature dependence of the centre shift, *CS*, which can be expressed as follows:

$$CS = IS(T) + SOD(T) \qquad (1)$$

where the first term, *IS*, represents the isomer shift, a quantity related to the charge-density of s-like electrons at the nucleus site. It was shown to be only weakly dependent on temperature [17]. The second term, *SOD*, the so-called second-order Doppler shift, is related to the atomic mean square displacement in the lattice, hence it is strongly temperature dependent. Assuming therefore that the total temperature dependence of *CS* goes via *SOD*, and using the Debye model to describe the lattice vibrations, the Debye temperature, $\Theta_D$, can be calculated by

$$CS(T) = IS(0) - \frac{3k_B T}{2mc}\left[\frac{3\Theta_D}{8T} + \left(\frac{T}{\Theta_D}\right)^3 \int_0^{\Theta_D/T} \frac{x^3}{e^x - 1}dx\right] \qquad (2)$$

where *m* is the mass of the $^{57}$Fe nucleus, $k_B$ the Boltzmann constant and *c* the velocity of light. Two examples for the temperature dependence of the average centre shift, *<CS>*, are shown in Fig. 5. The dependence of $\Theta_D$ on the vanadium concentration, *x*, is illustrated in Fig. 6 (top panel) from which it is clear that the $\Theta_D(x)$ behaviour is non-monotonous. Instead, two minima – one at $x \approx 40$ and the other at $x \approx 48$ – and a maximum at $x \approx 43$ are observed. Such behaviour and the fact that the amplitude of a change in $\Theta_D$ reaches ~40% is unexpected and it is quite puzzling. It is neither paralleled by the dependence of $<CS>(x)$ which, for the room temperature is illustrated in Fig. 7, nor by that of the Curie temperature or the average magnetic moment per Fe atom, both showing a monotonous decrease with $x$ [6]. To take into account a possible effect of the lattice parameters on $\Theta_D$, it was also presented as a function of the lattice constant, $a$ (middle panel), and of the unit cell volume, $V$ (bottom panel) in Fig. 6. As is evident, this procedure has not changed the character of $\Theta_D(x)$, except the value of $\Theta_D$ found for the nanocrystalline sample which, after the correction, has been shifted to a position close to the maximum observed around 43 at% V.

It seems therefore true that the anomaly revealed in the concentration dependence of $\Theta_D$ for the σ-FeV alloys have neither magnetic nor electronic nor geometric origin, and it challenges both theoretical calculations as well as further experimental studies including measurements of phonon spectra to get more detailed insite into the dynamics of the studied system. For comparison, the values of $\Theta_D$ found previously for the σ-FeCr alloys have been marked in Fig. 6 [8].

## 4. Conclusions

The results presented in this paper can be concluded as follows:

(a) Vanadium content, $x$, has a profound influence on lattice parameters and the Debye temperature in σ-FeV alloys.

(b) The lattice constants $a$ and $c$ increase linearly with $x$ at the rate characteristic of the lattice parameter. Consequently the volume of the unit cell for the most concentrated (59.0 at% V) sample is by 3.62% larger than the one for the least concentrated one (34.4 at% V).

(c) The Debye temperature shows a complex non-monotonic dependence on *x* which is not paralleled by that of the centre shift (charge-density), the Curie temperature and the magnetic moment.

**References**

[1] Hall E. O. and Algie S. H. 1966 Metall. Rev. **11** 61

[2] Read D. A. and Thomas E. H. 1966 IEEE Trans. Magn. MAG-**2** 415

[3] Read D. A., Thomas E. H. and Forsythe J. B. 1968 J. Phys. Chem. Solids **29** 1569

[4] Cieślak J., Reissner M., Steiner W. and Dubiel S. M. 2004 J. Magn. Magn. Mater. **272-276** 534

[5] Cieślak J, Reissner M., Steiner W. and Dubiel S. M.2008 Phys. Stat. Sol (a) 205 1794

[6] Cieślak J., Costa B. F. O., Dubiel S. M., Reissner M. and Steiner W. 2009 J. Magn. Magn. Mater. **321** 2160

[7] Cieślak J., Dubiel S. M., Żukrowski J., Reissner M. and Steiner W., Phys. Rev. B 2002 **65** 212301

[8] Cieślak J., Costa B. F. O., Dubiel S. M., Reissner M. and Steiner W. 2005 J. Phys.: Condens. Matter **17** 6889

[9] Kubaschewski O. 1982 in *Iron-Binary Phase Diagrams*, Springer Verlag, Berlin

[10] Blower R. and Cox G. J. 1970 J. Iron Steel Inst. August 769

[11] Rodriguez-Carjaval J. 1993 Physica B **192** 55

[12] Cieślak J., Reissner M., Dubiel S. M., Wernisch J. and Steiner W. 2008 J. Alloys Comp. **20** 20

[13] Dubiel S. M. and Zinn W. 1984 J. Magn. Magn. Mater. **45** 298

[14] Cieślak J., Toboła J., Dubiel S. M., Kaprzyk S., Steiner W. and Reissner M. 2008 J. Phys.: Condens. Matter **20** 235234

[15] Parwate D. V. and Garg A. N. 1984 J. Radionalitycal and Nucl. Chem. Lett. **87** 379

[16] Vincze I., Van Der Woude F. and Friedt J. M. 1986 Phys. Rev. B **33** 5050

[17] Wilgeroth S., Ulrich H. and Hesse J. 1984 J. Phys. F: Metal. Phys. **14** 387

**Figures**

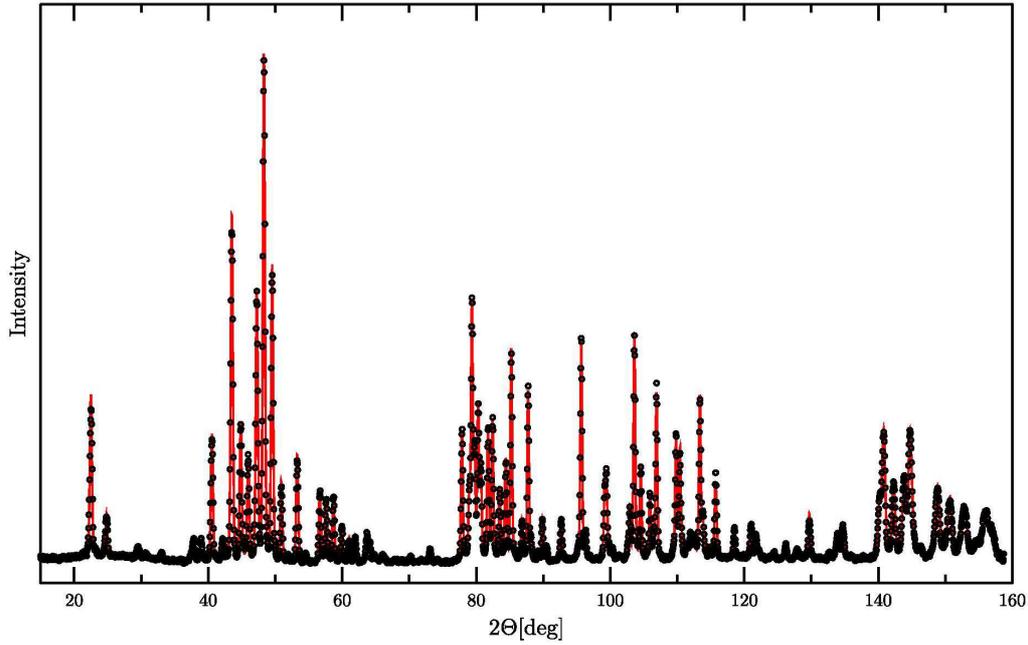

Fig. 1
Room temperature neutron diffraction pattern recorded on the σ-$Fe_{65.6}V_{34.4}$ sample together with the best-fit pattern in red.

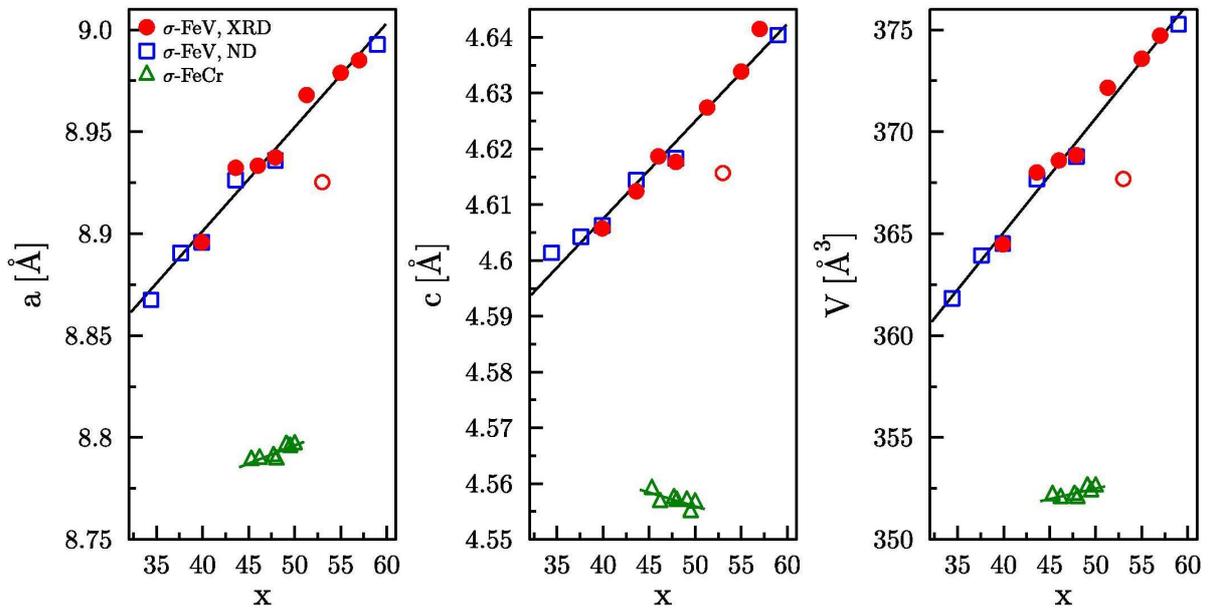

Fig. 2
Lattice parameters *a* and *c*, and the volume of the unite cell, *V*, as determined from the neutron (squares) and X-ray (circles) diffraction patterns for the investigated samples. Straight lines are only guides to the eyes. The point marked with open circle is for the nanocrystalline sample of σ-FeV. For comparison, the data found for the σ-phase in the Fe-Cr alloys are marked, too, with triangles.

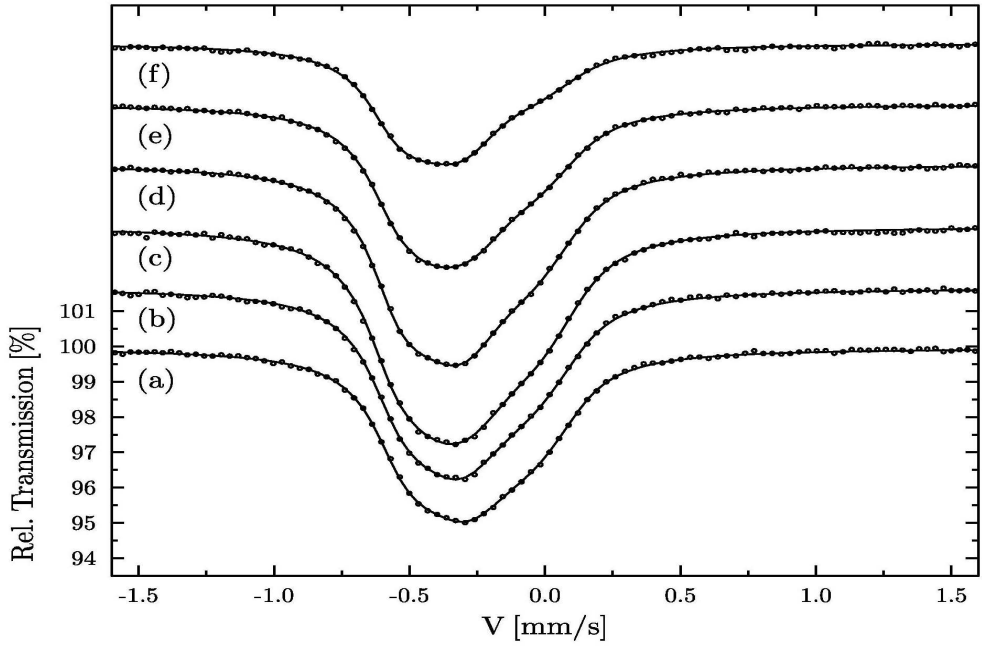

Fig. 3
Examples of the $^{57}$Fe Mössbauer spectra recorded at 295 K for the σ-$Fe_{100-x}V_x$ samples with $x$ equal to: (a) 55.0, (b) 47.9, (c) 46.0, (d) 43.6, (e) 42.5 and (f) 37.6. The best-fit spectra are indicated by full lines.

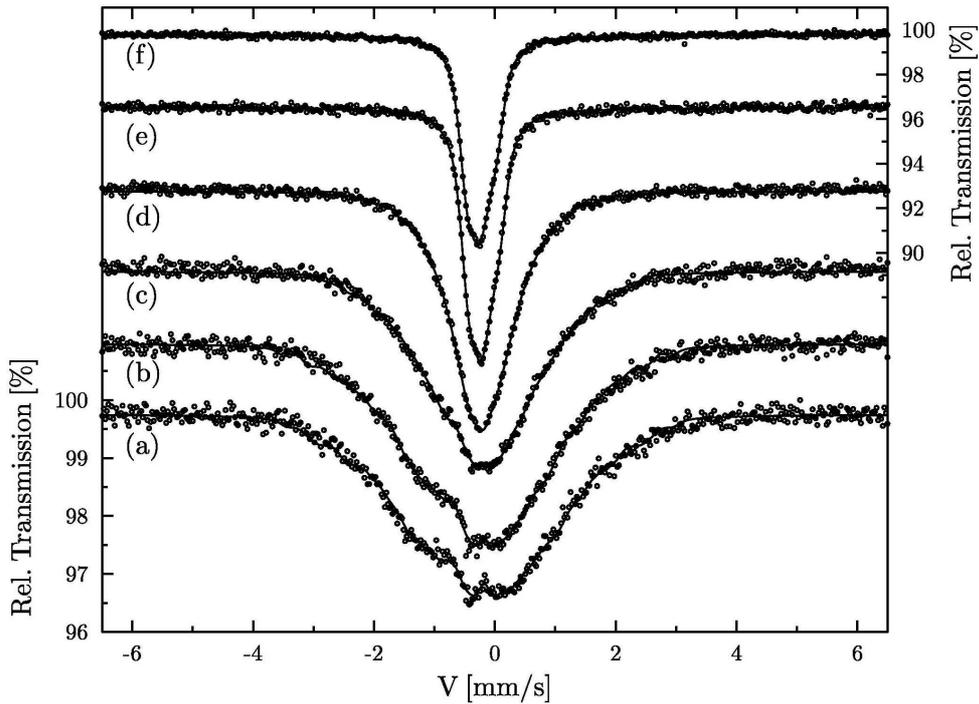

Fig. 4
$^{57}$Fe Mössbauer spectra recorded on the σ-$Fe_{60.1}V_{39.9}$ sample at (a) 4.2 K, (b) 50 K, (c) 100 K, (d) 150 K, (e) 200 K, and (f) 250 K. The best-fit spectra are indicated by full lines. The left-hand scale refers to spectrum (a) and the right-hand one to spectrum (f).

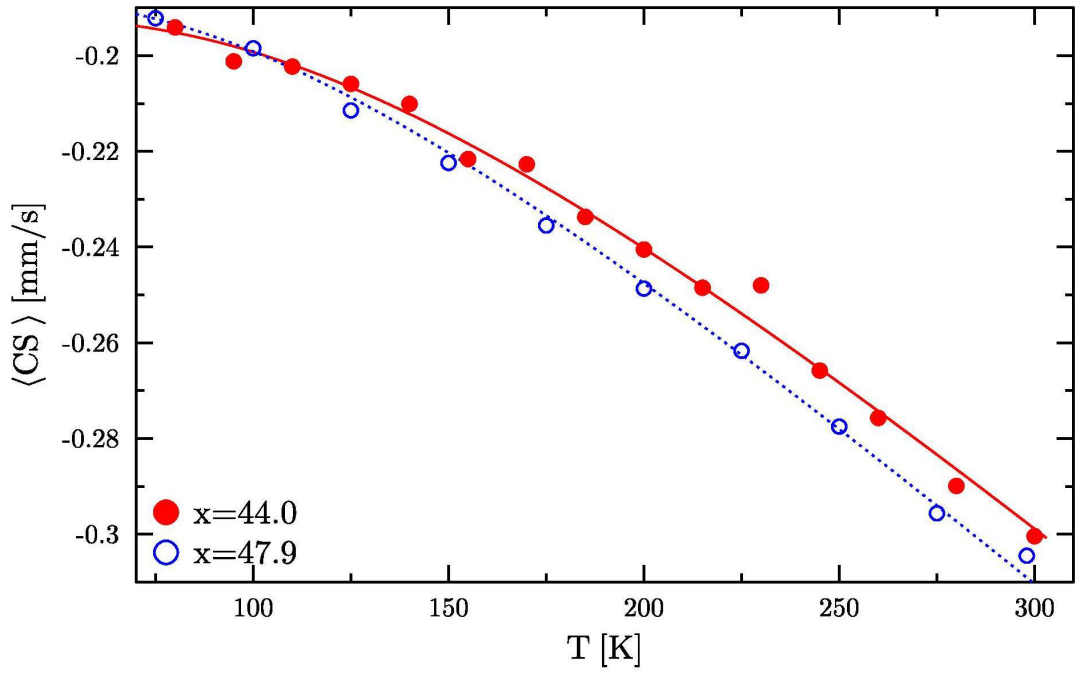

Fig. 5 Examples of the temperature dependence of the average centre shift, <*CS*>, for two σ-Fe$_{100-x}$V$_x$ samples with the *x* – value shown. The best fits to the experimental data in terms of equ. (2) are indicated by a full and a dotted line, respectively.

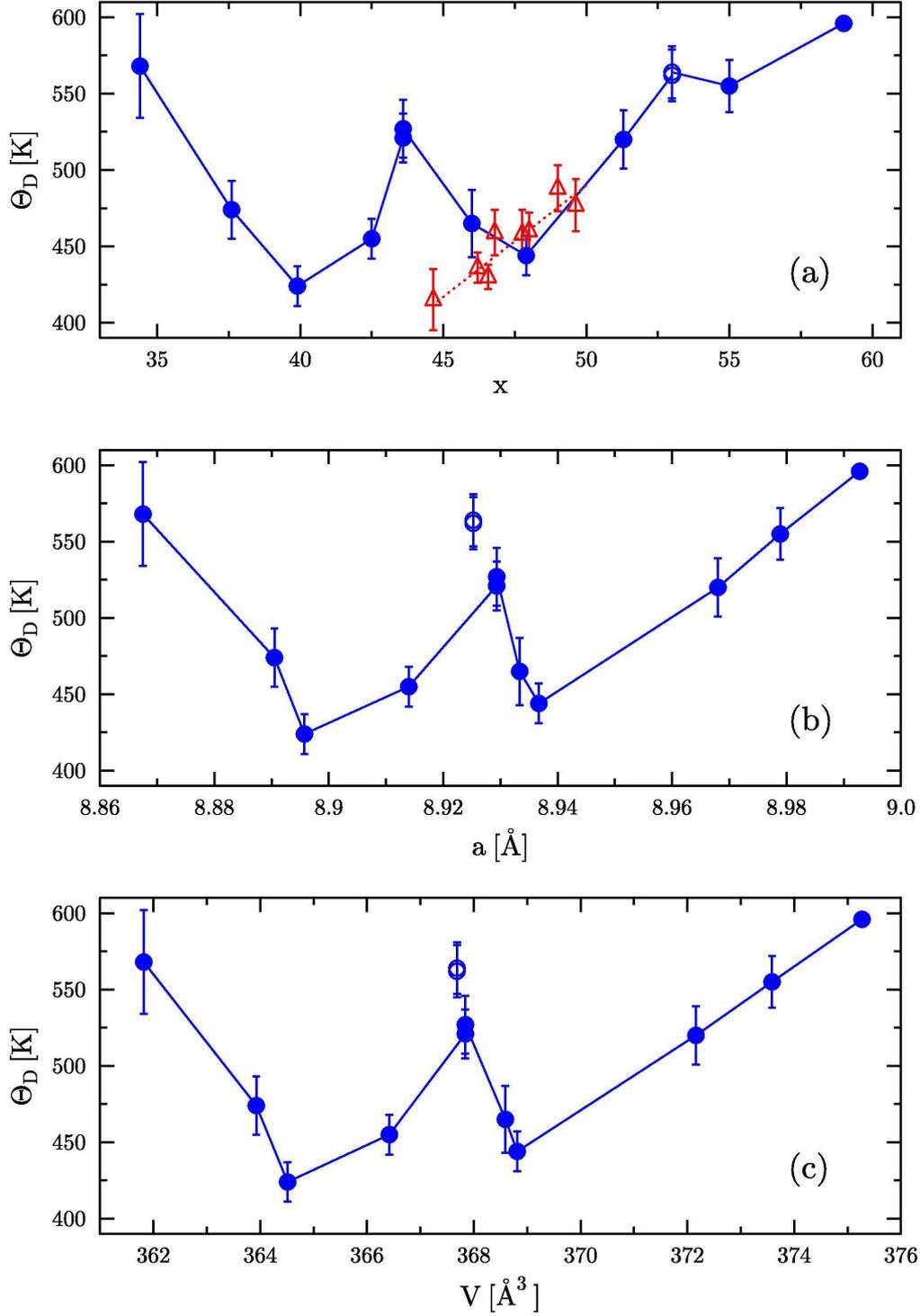

Fig. 6
Dependence of the Debye temperature, $\theta_D$, on vanadium concentration, $x$, (a) on the lattice constant, $a$ (b) and on the unit cell volume, $V$ (c) (full circles for the bulk samples and open circle for the nanocrystalline one) as determined in the present study, and on chromium concentration (triangles) as reported in Ref. 8. The solid and dotted lines are only guides to the eyes. The error bar in $x$ has a size of the symbol.

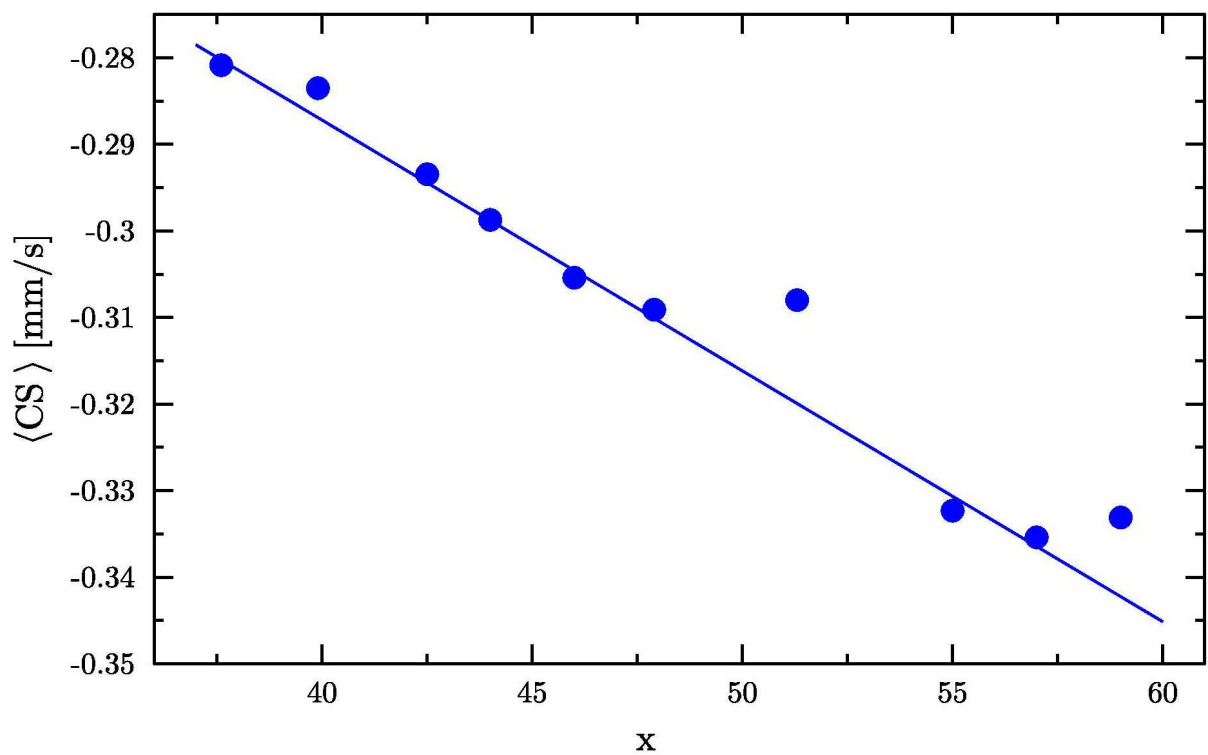

Fig. 7 Dependence of the average centre shift, $\langle CS \rangle$, determined from the spectra measured at room temperature on the content of vanadium, $x$. The solid line stands for the best linear fit to the data.